\documentclass[9pt,review]{livecoms}

\usepackage{lipsum} %
\usepackage[version=4]{mhchem}
\usepackage{siunitx}
\usepackage{booktabs}
\DeclareSIUnit\Molar{M}
\usepackage[italic]{mathastext}
\graphicspath{{figures/}}
\newcommand{\versionnumber}{1.3}  %
\newcommand{\githubrepository}{\url{https://github.com/ltalirz/livecoms-atomistic-software}}  %

\newcommand{\atsoft}{\href{https://atomistic.software}{atomistic.software}\ }
\newcommand{\atsoftgit}{\href{https://github.com/ltalirz/atomistic-software}{ltalirz/atomistic-software}\ }

\title{Trends in atomistic simulation software usage [\versionnumber]}

\author[1,2,3*]{Leopold Talirz}
\author[4]{Luca M. Ghiringhelli}
\author[1,3]{Berend Smit}
\affil[1]{Laboratory of Molecular Simulation (LSMO),
    Institut des Sciences et Ingenierie Chimiques,
    Valais, \'Ecole Polytechnique F\'ed\'erale de Lausanne,
    CH-1951 Sion, Switzerland}
\affil[2]{Theory and Simulation of Materials (THEOS),
    Facult\'e des Sciences et Techniques de l'Ing\'enieur,
    \'Ecole Polytechnique F\'ed\'erale de Lausanne,
    CH-1015 Lausanne, Switzerland}
\affil[3]{National Centre for Computational Design and Discovery
of Novel Materials (MARVEL), \'Ecole Polytechnique F\'ed\'erale de Lausanne,
CH-1015 Lausanne, Switzerland}
\affil[4]{The NOMAD Laboratory at the Fritz Haber Institute of the Max Planck Society and Humboldt University, Berlin, Germany}

\corr{leopold.talirz@gmail.com}{LT}  %

\orcid{Leopold Talirz}{0000-0002-1524-5903}
\orcid{Luca M. Ghiringhelli}{0000-0001-5099-3029}
\orcid{Berend Smit}{0000-0003-4653-8562}

\blurb{This LiveCoMS document is maintained online on GitHub at \githubrepository; to provide feedback, suggestions, or help improve it, please visit the GitHub repository and participate via the issue tracker.}

\pubDOI{10.XXXX/YYYYYYY}
\pubvolume{<volume>}
\pubissue{<issue>}
\pubyear{<year>}
\articlenum{<number>}
\datereceived{Day Month Year}
\dateaccepted{Day Month Year}

\begin{document}

\begin{frontmatter}
\maketitle

\begin{abstract}
Driven by the unprecedented computational power available to scientific research, the use of computers in solid-state physics, chemistry and materials science has been on a continuous rise.
This review focuses on the software used for the simulation of matter at the atomic scale.
We provide a comprehensive overview of major codes in the field, and analyze how citations to these codes in the academic literature have evolved since 2010.
An interactive version of the underlying data set is available at \href{https://atomistic.software}{https://atomistic.software}.
\end{abstract}

\end{frontmatter}

\section{Introduction}

\begin{figure}
    \includegraphics[width=0.5\textwidth]{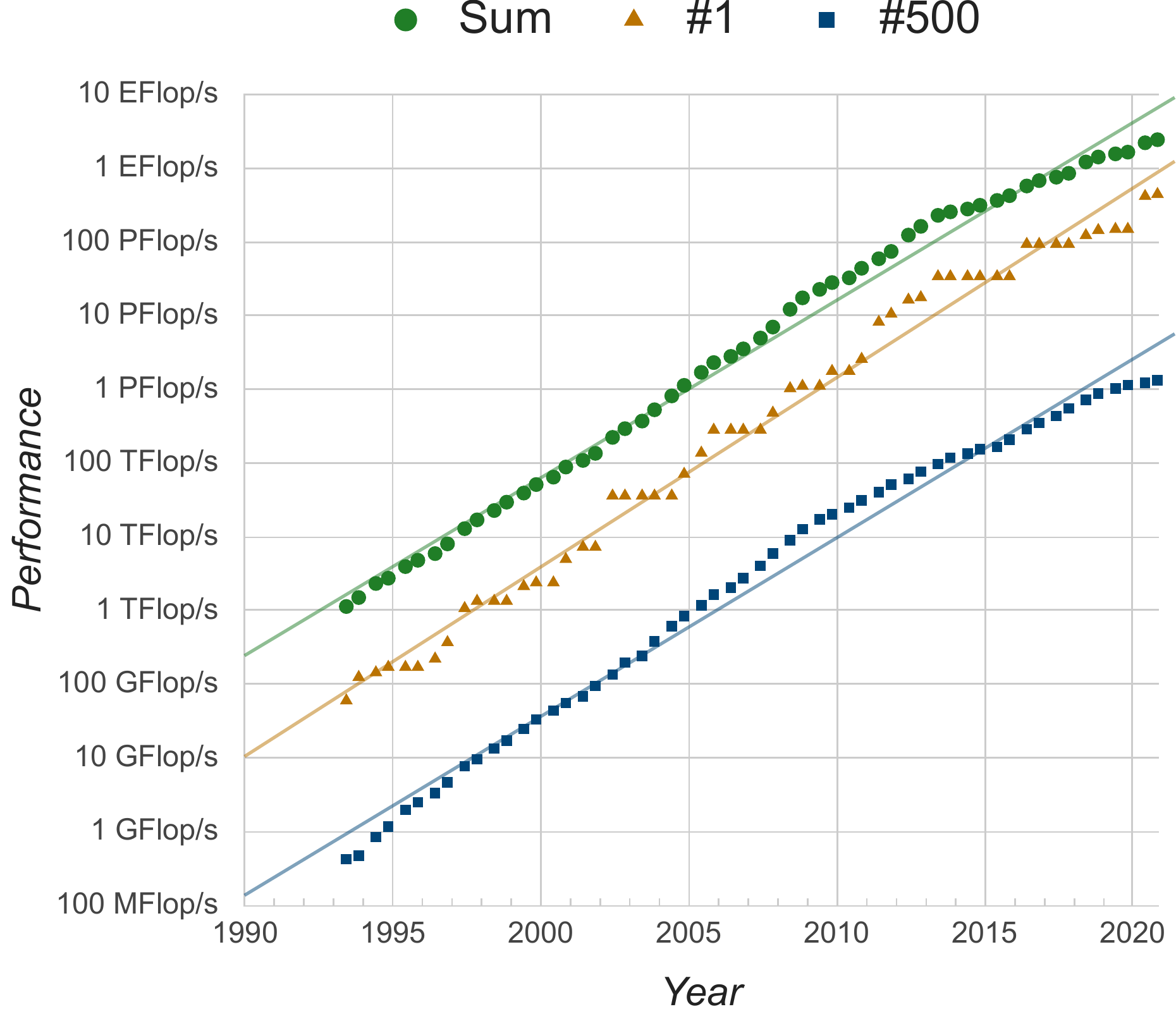}
    \caption{
        Performance of the top 500 supercomputers in the world from 1993 to 2020 in solving linear equations (Linpack benchmark), measured in 64bit floating point operations per second (FLOP/s).
        Shown are the sum of the entire list (green dots), the performance of the top machine (brown triangles), and performance of the bottom of the list (blue squares), together with least squares fits.
        Adapted from \href{https://www.top500.org/statistics/perfdevel/}{top500.org/statistics/perfdevel}.
    }
    \label{fig:top500}
\end{figure}

Scientists today have unprecedented access to computational power.
This statement would be unremarkable, were it not for the extent to which computational power has exploded.
Figure~\ref{fig:top500} shows the performance ranking of the top 500 supercomputers in the world over the last decades, including the performance of the top-ranked machine, the machine at the bottom of the list, and the sum of all 500.
Remarkably, the least-squares fit to the sum (green line) corresponds to a growth rate of ${\sim}75\%$ per year, which translates to a performance increase of more than one \emph{million} times over the last 25 years.
Similar improvements in commodity hardware mean that many of 2020's laptop computers would have made the top 500 list of the early 2000s \cite{Lehtola2021}.

First versions of quantum-chemistry codes, such as Gaussian \cite{Hehre1970}, 
were already released in the 1970s,
followed by force-field codes, such as GROMOS \cite{VANGUNSTEREN1982}, 
and periodic density-functional theory (DFT) codes, such as CASTEP \cite{Payne1992a}, in the 1980s and 1990s.
In other words, many of these atomistic simulation engines have been around during this explosion of computational power, continuously evolving to take advantage of new algorithms, processor architectures, increasing parallelism and, more recently, dedicated accelerator hardware.
Over time, they have developed from instruments for specialists to proven and tested tools in the arsenal of practitioners in physics, chemistry, and materials science.

\begin{table*}
    \centering

\begin{tabular}{lllS[table-format=5.0]}
    \toprule
    Title                                                                 & Authors                    & Journal    & {\# cited} \\ \midrule
{Generalized gradient approximation made simple} & Perdew, Burke, Ernzerhof   & PRL (1996) & 108099  \\
{Development of the Colle-Salvetti correlation-energy formula … }       & Lee, Yang, Parr            & PRB (1988) & 77473   \\
{Efficient iterative schemes for ab initio total-energy calculations …} & Kresse, Furthmüller        & PRB (1996) & 58176   \\
{Projector augmented-wave method                                      } & Blöchl                     & PRB (1994) & 43455   \\
{Self-consistent equations including exchange and correlation …       } & Kohn, Sham                 & PR (1965)  & 42795   \\
{From ultrasoft pseudopotentials to the projector augmented-wave …    } & Kresse, Joubert            & PRB (1999) & 42485   \\
{Special points for Brillouin-zone Integrations                       } & Monkhorst, Pack            & PRB (1976) & 41232   \\
{Density-functional exchange-energy approximation with correct …      } & Becke                      & PRA (1988) & 41142   \\
{Inhomogeneous electron gas                                           } & Hohenberg, Kohn            & PRB (1964) & 35445   \\
{Ab initio molecular dynamics for liquid metals                       } & Kresse, Hafner             & PRB (1993) & 23192   \\
\end{tabular}

    \caption{
        Top ten most highly cited articles published by the American Physical Society,
        all of which deal with density functional theory and its practical application.
        Data collected from the Web of Science on June 16th, 2021.
    }
    \label{tab:aps-most-cited}
\end{table*}

Records of the pervasive use of these tools can be found in the scientific literature.
In a 2014 survey, van Noorden et al.\ found that 12 of the top 100 most cited papers of all time were on density-functional theory \cite{VanNoorden2014}.
As with other examples in van Noorden's list, the flood of citations are indicative of papers being cited by the many practitioners (here: of density-functional theory) rather than the few method developers.
If one focuses the analysis on articles published in physics journals, the footprint of density-functional theory grows even further:
For example, table~\ref{tab:aps-most-cited} shows the top ten most cited papers published by journals of the American Physical Society.
All of them are related to density-functional theory and its application.

\begin{table*}
    \centering
    \begin{tabular}{llllS[table-format=7.0]} \toprule
        Software & Version & Main language& {Contributors 2020} & {Lines of code} \\ \midrule
        NWChem \cite{Apra2020}&7.0.2&Fortran& 11 &5034139\textsuperscript{*} \\
        LAMMPS \cite{Plimpton1995}&27May2021&C++& 51 & 1039872  \\
        CP2K \cite{Kuhne2020a}&8.2&Fortran & 38 & 911367 \\
        ABINIT \cite{Gonze2020}&9.4.2&Fortran& 21 & 719048 \\
Quantum ESPRESSO \cite{Giannozzi2020}&6.7.0&Fortran& 43& 604666 \\
GROMACS \cite{Abraham2015}&2021.2&C++& 25 & 589108 \\
Psi4 \cite{Smith2020}&1.3.2&C++& 34 & 496858 \\
SIESTA \cite{Garcia2020}&5.0.0-alpha&Fortran& 4 & 411829 \\
Octopus \cite{Tancogne-Dejean2020}&10.5&Fortran& 10 & 297825 \\
OpenMM \cite{Eastman2017}&7.5.1&C++& 21 & 256452 \\
xtb \cite{Bannwarth2021}&6.4.1&Fortran& 18 & 124344 \\\hline
Linux kernel&5.13&C& >4000 \cite{LinuxHistory2020} & 20904410\\
    \end{tabular}\\
    \caption{
        Counting lines of code for 11 popular open-source atomistic simulation engines (and the Linux kernel for comparison), using the latest releases as of June 2021.
        Line counts are determined by cloc v1.6.0 \cite{Danial2021} and exclude blank lines, comments, and markup languages (detailed reports in the supporting information).
        Contributors for the year 2020 were determined by counting the number of different committers to the source code from January 1st 2020 to January 1st 2021 (numbers for the Linux kernel are from 2019).
        \textsuperscript{*} Roughly 3 million lines of code of NWChem are computer-generated.
    }
    \label{tab:lines-of-code}
\end{table*}

There used to be a time when it was commonplace for computational condensed-matter physicists and quantum chemists to write their own electronic-structure code, and many of the atomistic-simulation engines that are in broad use today have started this way.
Over the years, however, many of these engines have developed into complex software distributions.
Table \ref{tab:lines-of-code} shows counts for the lines of code in some of today's popular open-source simulation engines: they range from hundreds of thousands to millions of lines of code, typically written in Fortran or C++, with similar numbers being reported for commercial packages \cite{Krylov2015}.
While statistics like these are by no means accurate measures of code complexity (and developers follow different approaches to packaging and outsourcing of functionality to external libraries), they nevertheless suggest that many of these code bases are too large to be sustained by any single person.

This poses important questions for how to sustain these software projects going forward: questions of funding, business models, and software licenses.
Proponents of the open-source route argue that it democratizes research and education by removing barriers for both users and developers, and that science carried out with commercial software is harder to verify and reproduce \cite{Stodden2009,Gezelter2015,Lehtola2021}.
The open-source model can also be adopted irrespective of the size of a code's target user group, while commercial activity tends to require a minimal market size.
On the other hand, the strong focus on innovation and development often found in open-source scientific software can negatively impact usability and quality of documentation \cite{Swarts2019}.
Proponents of commercial licenses argue that making academic software accessible to a broader community is a technical task best left to professional software engineers, and that the resulting gain in scientific productivity can easily outweigh the license fees that pay for it \cite{Krylov2015}.
We note that open-source and commercial activity do not necessarily exclude each other: numerous software companies are built around an open-source core, and we start seeing first examples in atomistic modelling as well (e.g. Molcas/OpenMolcas \cite{Fdez.Galvan2019}).
Overall, it has been suggested that ``scientific publications are a more sound metric [of the scientific impact of software] than either the price of a product or whether its source code is available in the public domain'' \cite{Krylov2015}.
Providing a peek into this citation record is one of the reasons for creating the \atsoft collection.

The other reason is a practical one: 
When young scientists start their first research project in atomistic simulations, they often have no grasp of the extent of this software ecosystem, let alone of current trends in the field -- at least this is the personal experience of the authors of this review.
Software choices in research groups are therefore often informed by what other members of the group already use.
This makes sense: colleagues have vetted the code for the type of problems the research group is working on, and built up expertise around which of the many knobs to turn in order to find the sweet spot between efficiency and accuracy.

\begin{figure}
    \includegraphics[width=0.5\textwidth]{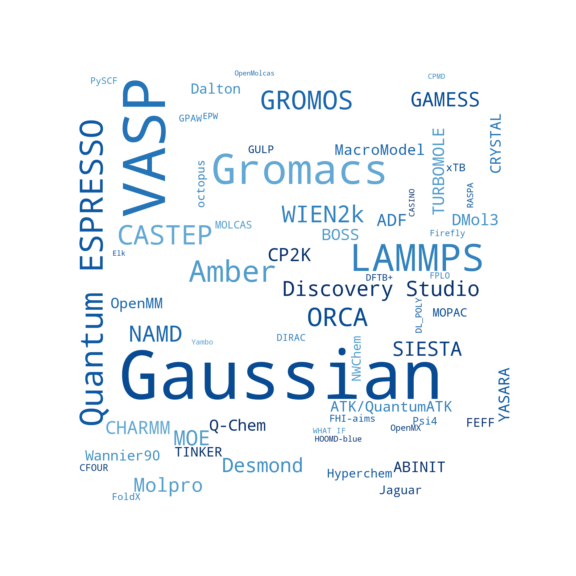}
    \caption{
        Highly cited atomistic simulation engines in the scientific literature. 
        Font size scaled (approximately) by the number of citations during the year 2020 as reported by Google Scholar.
    }
    \label{fig:wordcloud}
\end{figure}

But what if that code is no longer actively developed?
What if there was another code that was better suited to solve the specific research problem at hand? 
That had a larger user/developer community? 
Was free instead of commercial? 
Was open source instead of closed source? 
The goal of the \atsoft collection is to provide a comprehensive overview of all major atomistic simulation engines (cf. Figure~\ref{fig:wordcloud}), and to help newcomers to the field as well as experienced practitioners and software developers find better answers to some of these questions.

Readers interested mainly in the results are invited to go straight to the \href{atomistic.software}{atomistic.software} web site that displays the data discussed in this review.
For those wanting to know more, the following sections provide details on the methodologies used, and discuss some of the trends that can be observed.

\section{The atomistic.software collection}

\subsection{Overview}

The \atsoft collection draws upon existing lists of atomistic simulation codes \cite{qm-codes-wiki,mm-codes-wiki,sklog-wiki,Pirhadi2016,MolSSI2021}, in particular the "Major codes in electronic-structure theory, quantum chemistry, and molecular-dynamics" \cite{Ghiringhelli2017} maintained by the NOMAD Centre of Excellence from 2017-2019. 
It enriches these with annual citation data from the Google Scholar search engine, which provides an overview of the current usage landscape as well as ongoing trends, both at the level of individual codes and at the ecosystem level.

\begin{figure}
    \includegraphics[width=0.5\textwidth]{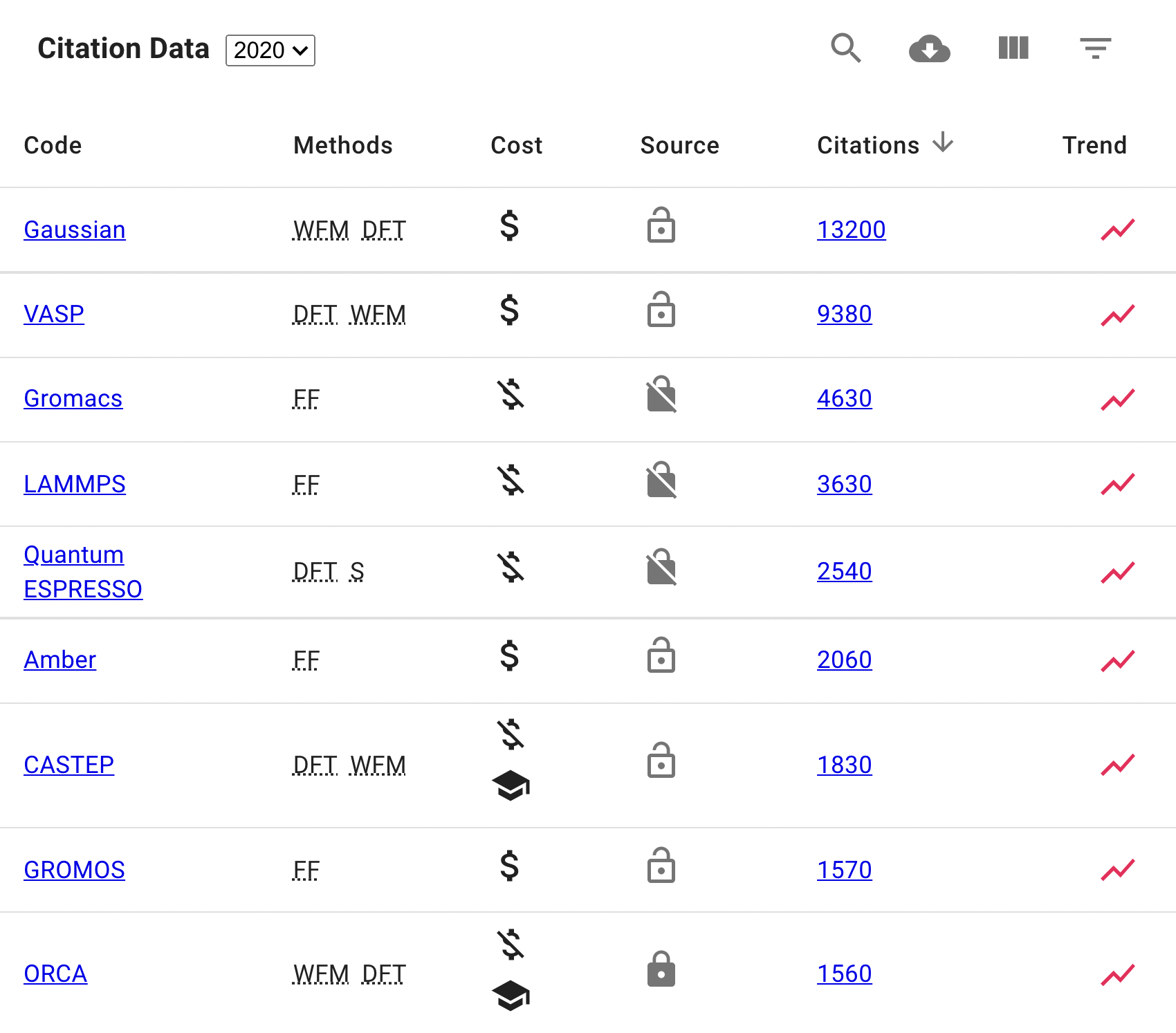}
    \caption{
        Overview table of atomistic simulation engines, sorted by how often they are referenced on Google Scholar during the previous year (here: 2020).
        A drop-down menu provides access to annual citation data reaching back to the year 2010.
    }
    \label{fig:overview}

\end{figure}
\begin{figure}
    \includegraphics[width=0.5\textwidth]{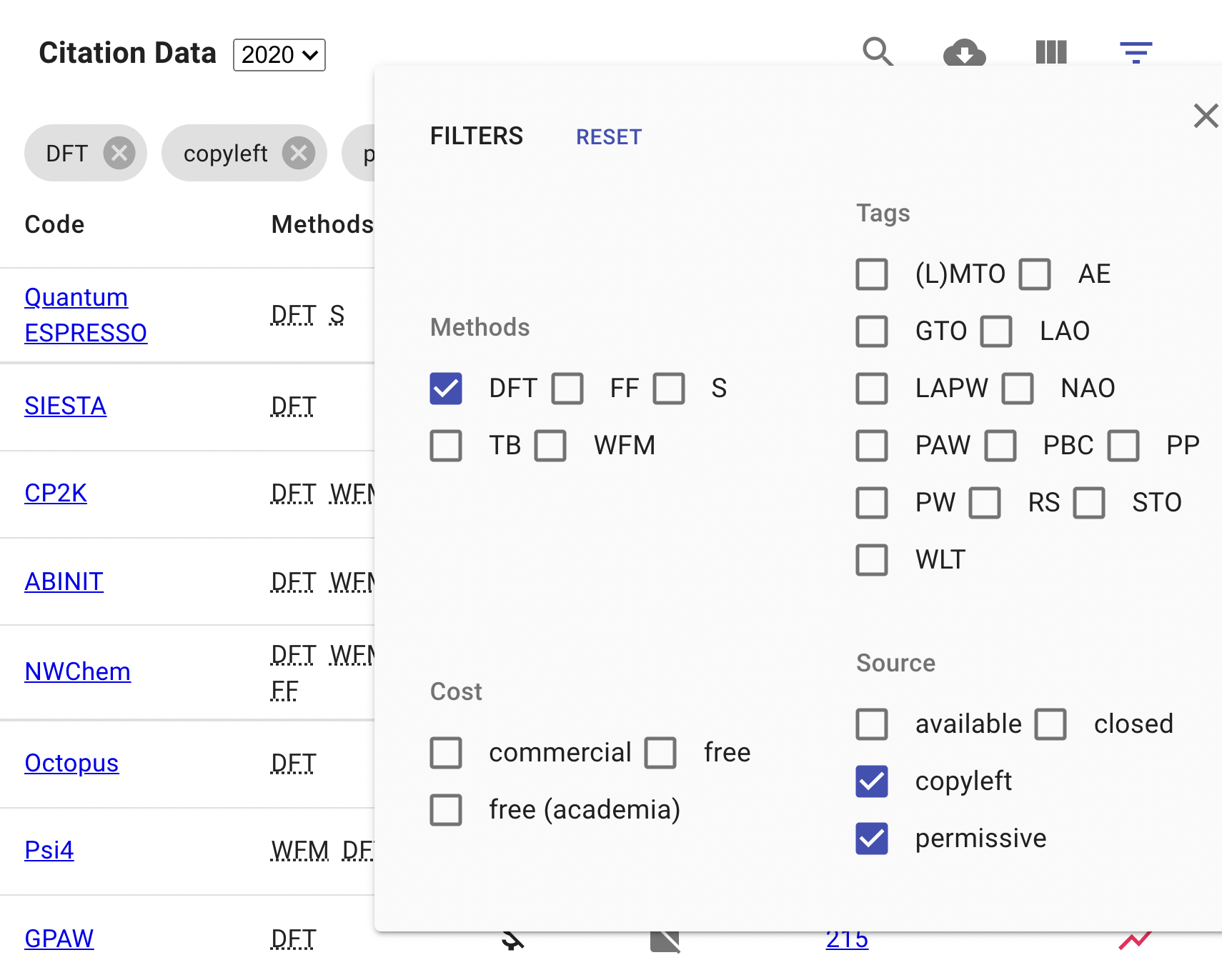}
    \caption{
        Filtering for density-functional theory codes with both permissive (P) and copyleft (CL) open-source licenses.
    }
    \label{fig:overview-filter}
\end{figure}

The overview table Fig.~\ref{fig:overview} lists all codes in the data set, ordered by how often they are referenced by articles indexed in Google Scholar. 
Clicking on the citation count opens the corresponding query on Google Scholar, so users can sift through the references one by one and discover what science is being done with this software.
Clicking on the name of the code instead opens the code's homepage for further information.

Users can filter codes by the methods or basis sets they use, and select only codes that are commercial, free or open-source (Fig.~\ref{fig:overview-filter}). 
Hovering with the mouse over any abbreviation in the list opens a tool-tip with an explanation.
Further metadata include information on which range of the periodic table the code covers, 
available installation routes, 
support for parallelization/acceleration,
support for standard APIs,
and the  availability of benchmarks.
In order not to clutter the interface, not all metadata is displayed by default but columns can be added/removed via the "View Columns" button.

\begin{figure}
    \includegraphics[width=0.5\textwidth]{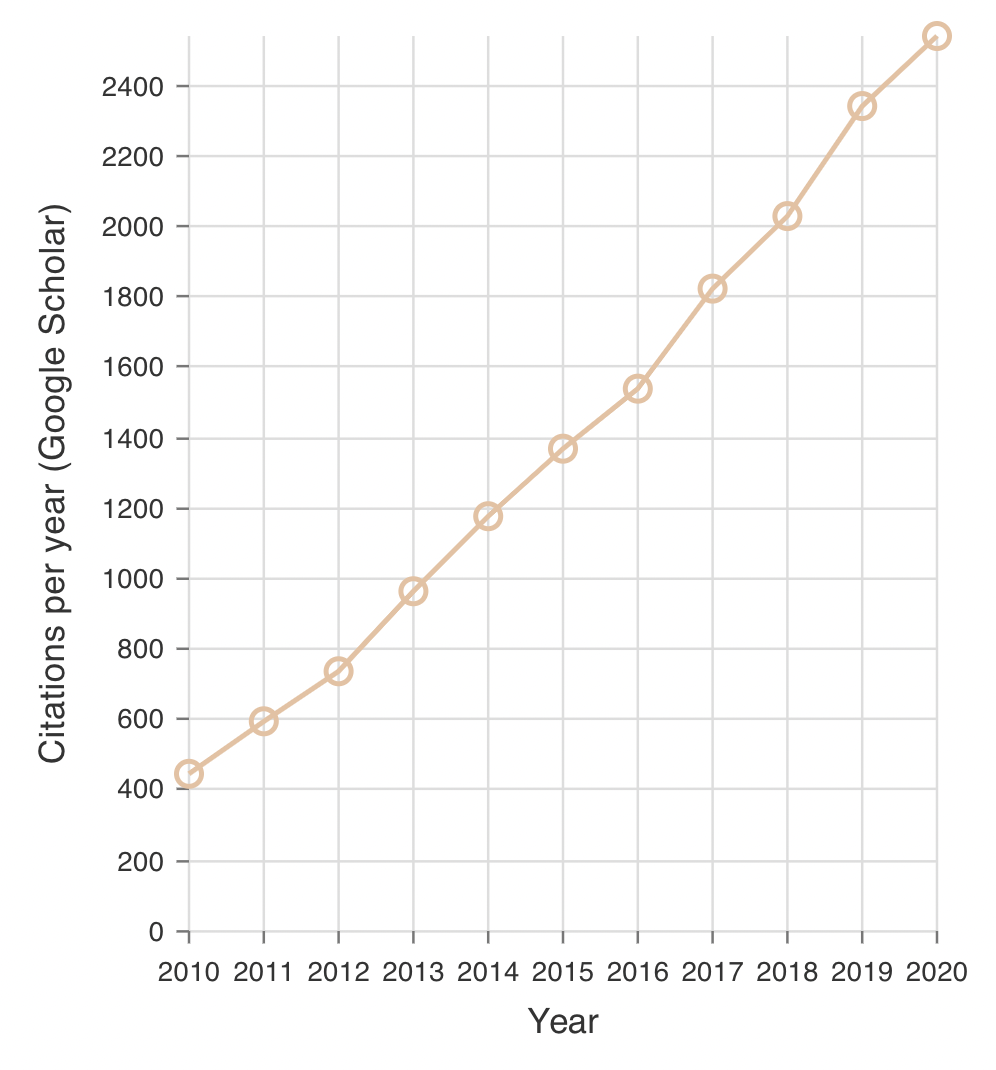}
    \caption{
        Citation trend for the Quantum ESPRESSO code.
    }
    \label{fig:quantum-espresso}
\end{figure}

Besides the generic overview, each engine comes with a citation "trend" over the last couple of years, which serves as an indicator of how its user community has developed over time (Figure~\ref{fig:quantum-espresso}).

Finally, the statistics page looks at the top codes by citation growth, indicating a rapidly growing user community. 
Ranking by absolute growth naturally favors established codes, while considering relative growth provides insight into the dynamics of new contenders in the list.
Ideally, codes rank highly in both metrics as is currently the case, e.g., for Desmond \cite{Desmond2021} and OpenMM \cite{Eastman2017}.

\begin{figure}
    \includegraphics[width=0.5\textwidth]{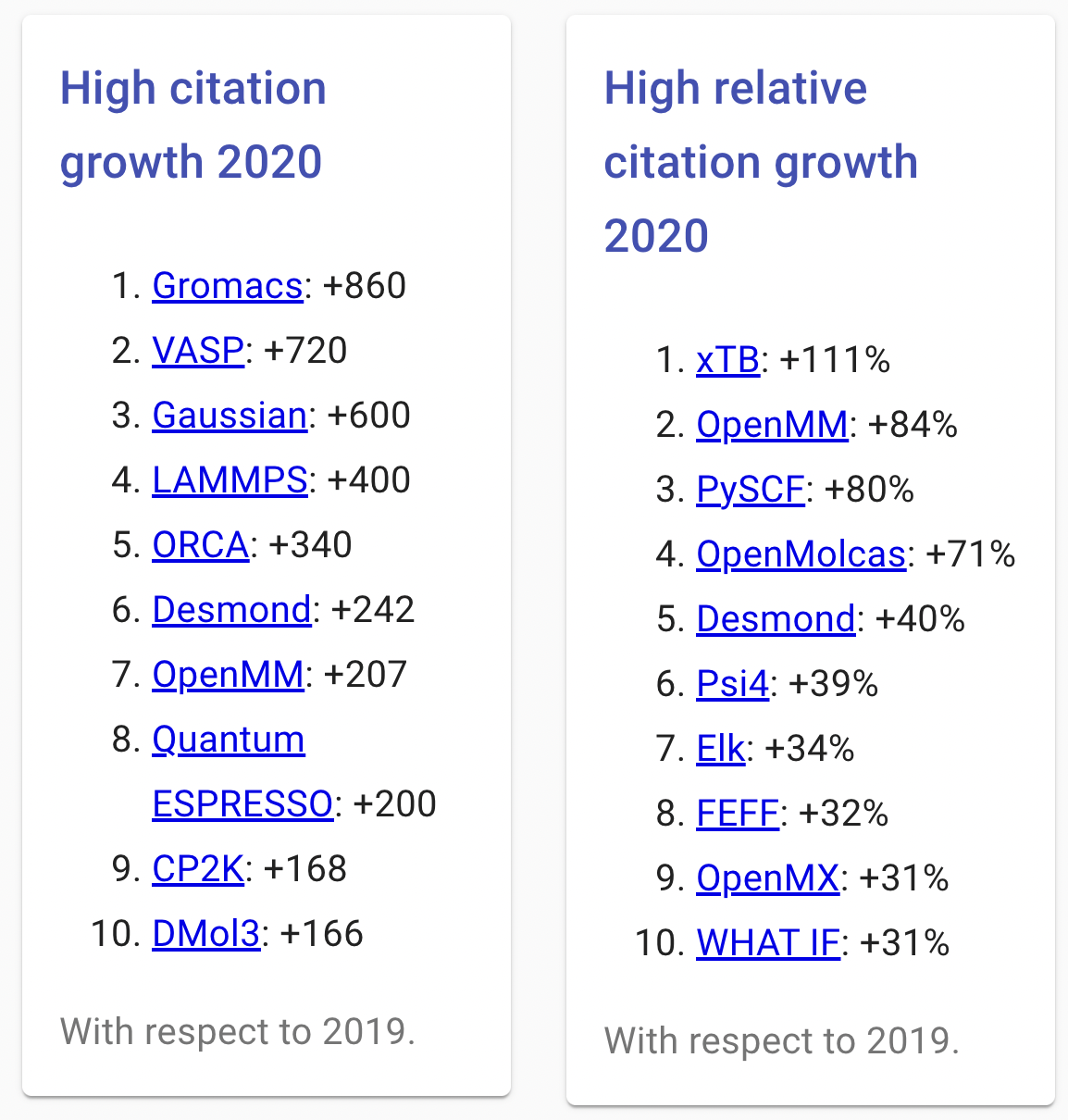}
    \caption{
        2020 rankings for absolute and relative citation growth with respect to the previous year.
    }
    \label{fig:highly-cited}
\end{figure}

A word of caution: The popularity of a code is a factor of many variables (starting, e.g., with the size of the target audience) -- please do not choose the code for your next research project merely based on its ranking in this list.
\atsoft links to the scholar query for the papers citing the code, thus making it quick and easy to get an impression of the research the code is currently used for.
Yet, certain aspects of a code are likely to correlate with popularity, such as how much used/tested the software is; how many Q\&A resources one is likely to find online or how much tooling there is likely around this code, etc. 
From a software-developer perspective, knowing the popularity of a code can be useful for gauging the potential impact of supporting the code in your own tool (such as a workflow manager, visualization software, \ldots).
And, finally, the citation trend provides an interesting peek into the future -- is the user community growing, stagnating or decreasing?

\subsection{Scope}

\atsoft uses the following working definition of an atomistic simulation engine:

\begin{quote}
A piece of software that, given two sets of atomic elements and positions (and, possibly, bond network), can compute their relative internal energies. 
In almost all cases, engines will also be able to compute the derivative of the energy with respect to the positions, i.e. the forces on the atoms, and thus be able to perform tasks like geometry optimizations or molecular dynamics.
\end{quote}

This covers the Density-Functional Theory (\verb|DFT|), Wave-Function Methods (\verb|WFM|), Quantum Monte Carlo (\verb|QMC|), Tight-Binding (\verb|TB|), and Force-Field (\verb|FF|) categories.
Codes in the Spectroscopy (\verb|S|) category are not necessarily simulation engines in the above sense, but compute the response of a given atomic structure to an external excitation (via photons, electrons, \ldots).

\atsoft aims to be a \emph{comprehensive} list of all major atomistic simulation engines, with annual updates going forward. 
Since there is a long tail of simulation engines with a limited user base, a relevance criterion is introduced in order to keep maintenance of the list manageable.
The criterion has been set to having at least one year with 100 citations or more.
The value of 100 is not set in stone and could be re-evaluated in the future, once the list has had some time to consolidate.
A "watch list" is kept of codes that do not yet meet the criterion.

\label{sec:methodology}
\subsection{Methodology \& Limitations}

Approximate citation counts are obtained from Google Scholar as follows:
\begin{enumerate}
    \item Search for name of the code and the last name of a representative developer who is a coauthor of all key publications on the software (vast majority of codes). 
        If no such coauthor exists, this is easily extended to searching for the presence of one of multiple author names (or a company name for commercial codes).
    \item When the name of the code is too common a search term, additional search terms may be added or citations of a major reference article are counted (minority of codes)
\end{enumerate}
Google Scholar was chosen over alternative sources like the Web of Science or Scopus, since it provides full-text search and is available for free, thus enabling direct links to the queries.
The supporting information contains a case study comparing citation counts from the search-based Google Scholar approach against counting the citations of reference papers (both in Google Scholar and the Web of Science).

Owing to the lack of standardization in today's software citation practices \cite{vandeSandt2019a,vandeSandt2020}, the citation counts reported here are necessarily approximate.
Shortcomings include the following:
\begin{itemize}
    \item While spot checks have been performed to weed out false matches (and reports on the \atsoftgit GitHub repository are highly welcome), details of the query can have significant impact on the number of results.
        This means, in particular, that the ranking by absolute number of citations is not set in stone and may be subject to change if more accurate search terms are identified. 
    \item Citation counts reported by Google Scholar are not entirely static, even for years that lie in the past. 
Reasons may include new publishers being indexed, more text being extracted, different citations being disambiguated, or even the heuristic evolving that predicts the total number of results. 
In our experience, citation data for the previous year can be subject to significant (upwards) fluctuation, while citation data for years further in the past are quite stable.
For this reason, for each data point the date of collection is recorded in the source code repository.
    \item Counting citations does not directly measure how often simulation codes are \emph{used} but how often they are \emph{referenced} in the scientific literature.
This may involve some systematic bias, for example if popular codes are more likely to be mentioned without being used,
or if a software targets industrial users who may be less likely to publish their results.
\end{itemize}

The caveats listed above mainly affect the \emph{absolute} number of citations reported, and thus the ranking of codes.
Citation trends on the individual code level should be more robust, and potential shortcomings in that domain (e.g.\ missing citations to a new reference paper with different authors) can be addressed by adapting the corresponding query.

The categorization of codes in terms of methods, tags and licenses is an evolution of the classification devised by the NOMAD list \cite{Ghiringhelli2017}.
For the sake of this data set, the following terminology has been adopted:
\begin{itemize}
    \item \emph{commercial}: payment required to obtain the software%
\footnote{This definition of \emph{commercial} does not necessarily mean \emph{for profit} -- 
research groups may price software below the actual development costs.
Furthermore, commercial licenses may exempt specific groups from payment, e.g. based on country of residence, membership in consortia, etc.
}
    \item \emph{free for academic use}: free for academics around the world%
\footnote{Some academic licenses may explicitly exclude researchers from specific countries.}
    \item \emph{free}: free to use for anyone, possibly after registration
    \item \emph{source available}: source code available either for free or against payment
    \item \emph{open-source}: open-source license approved by the Open Source Initiative (OSI, \url{https://opensource.org/})
\end{itemize}

We note that license terms can (and sometimes do) change over time. 
This is currently not reflected in this data set (only the latest license terms are recorded), but could be taken into account in future updates.

All data, as well as the source code of the web application running on \atsoft are hosted in the \atsoftgit GitHub repository.
The data is released under version 4 of the Creative Commons Share-Alike Attribution International License (CC-BY-SA).
The web application is written in JavaScript using the React framework (\url{reactjs.org}) and released under version 3 of the Affero General Public License (AGPL).

\subsection{Trends}

Extensive cross-checks of \atsoft against other lists \cite{qm-codes-wiki,mm-codes-wiki,sklog-wiki,Pirhadi2016,MolSSI2021} 
suggest that the collection is already fairly complete,
and can thus enable a look at the landscape of atomistic simulation software as a whole. 
Today's atomistic simulation engines are highly sophisticated pieces of software that each take many human-years of development, and developers have chosen different routes to support these efforts: 
from commercial to free, from closed source to open, and many shades of grey in between.
One question we can ask is: How do commercial codes fare versus their free competitors?

\begin{figure}
    \includegraphics[width=0.5\textwidth]{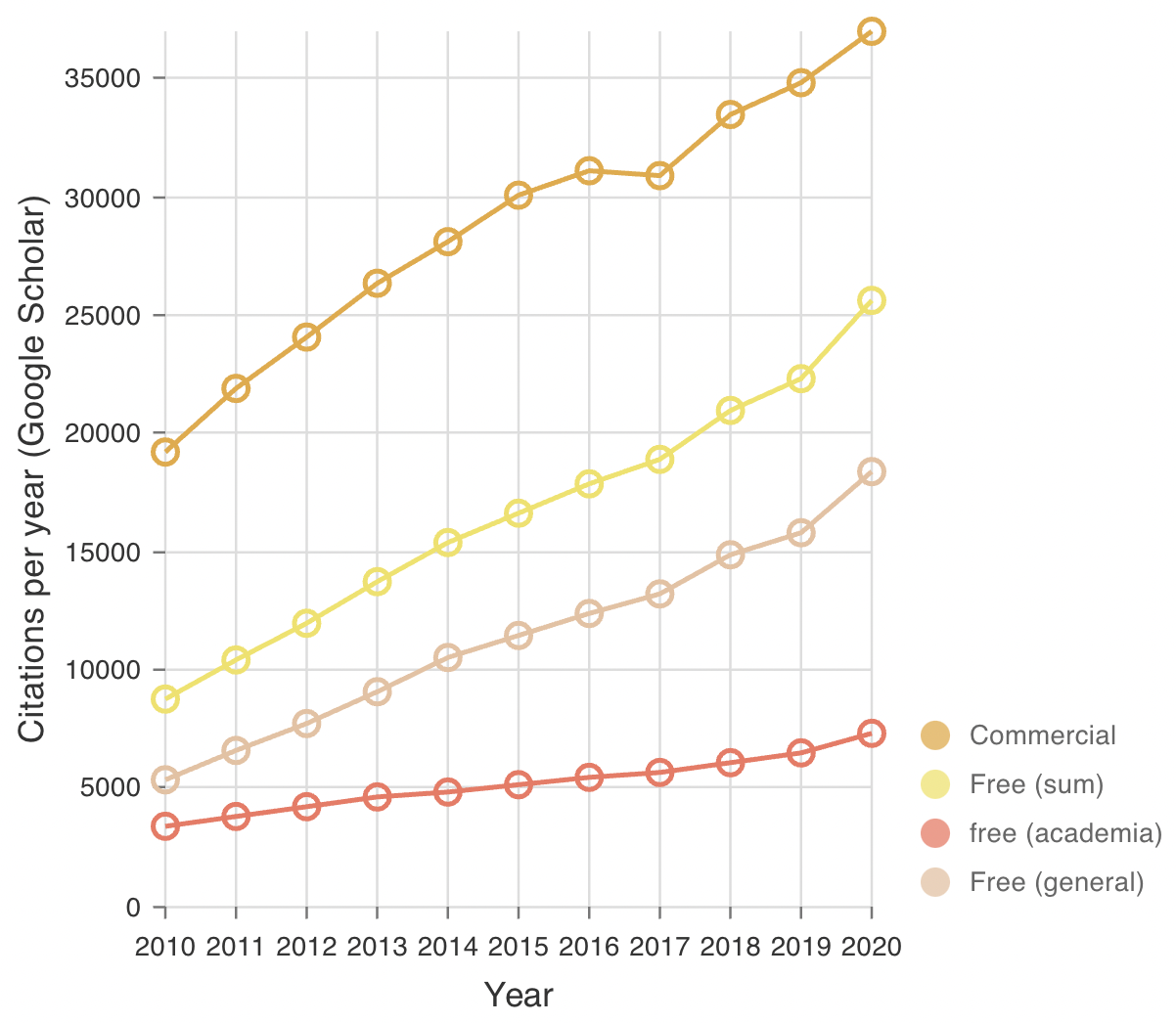}
    \caption{
        Citations of commercial vs free codes, including a breakdown into those free for general use and those free for academic use only.
    }
    \label{fig:money}
\end{figure}

Figure~\ref{fig:money} compares the compound citations to commercial and free codes. 
It illustrates that commercial codes are alive and well:
they are ahead in terms of citations gathered, and have been ahead throughout the last decade, with Gaussian \cite{g16} and VASP \cite{Kresse1996a} together accounting for more than half of all citations of the 23 commercial codes. 
At the same time, citations of the 40 free codes (including those that are only free for academic use) have been growing roughly at the same \emph{absolute} rate, mostly driven by the codes that are free for general use.

We note some caveats that apply to this statistic:
\begin{itemize}
    \item 
        The current dataset only records the \emph{latest} license conditions, while some codes (e.g. CASTEP \cite{Clark2005} or Dalton \cite{Aidas2014}) have moved to more open license terms over time, thus switching categories.
    \item For codes that are free for academic use only, some researchers may prefer to use the commercial version (e.g. using CASTEP through Biovia's Materials Studio software \cite{Meunier2021}).
\end{itemize}
It seems safe to conclude, however, that -- while commercial codes remain highly popular -- free codes are slowly gaining market share.

\begin{figure}
    \includegraphics[width=0.5\textwidth]{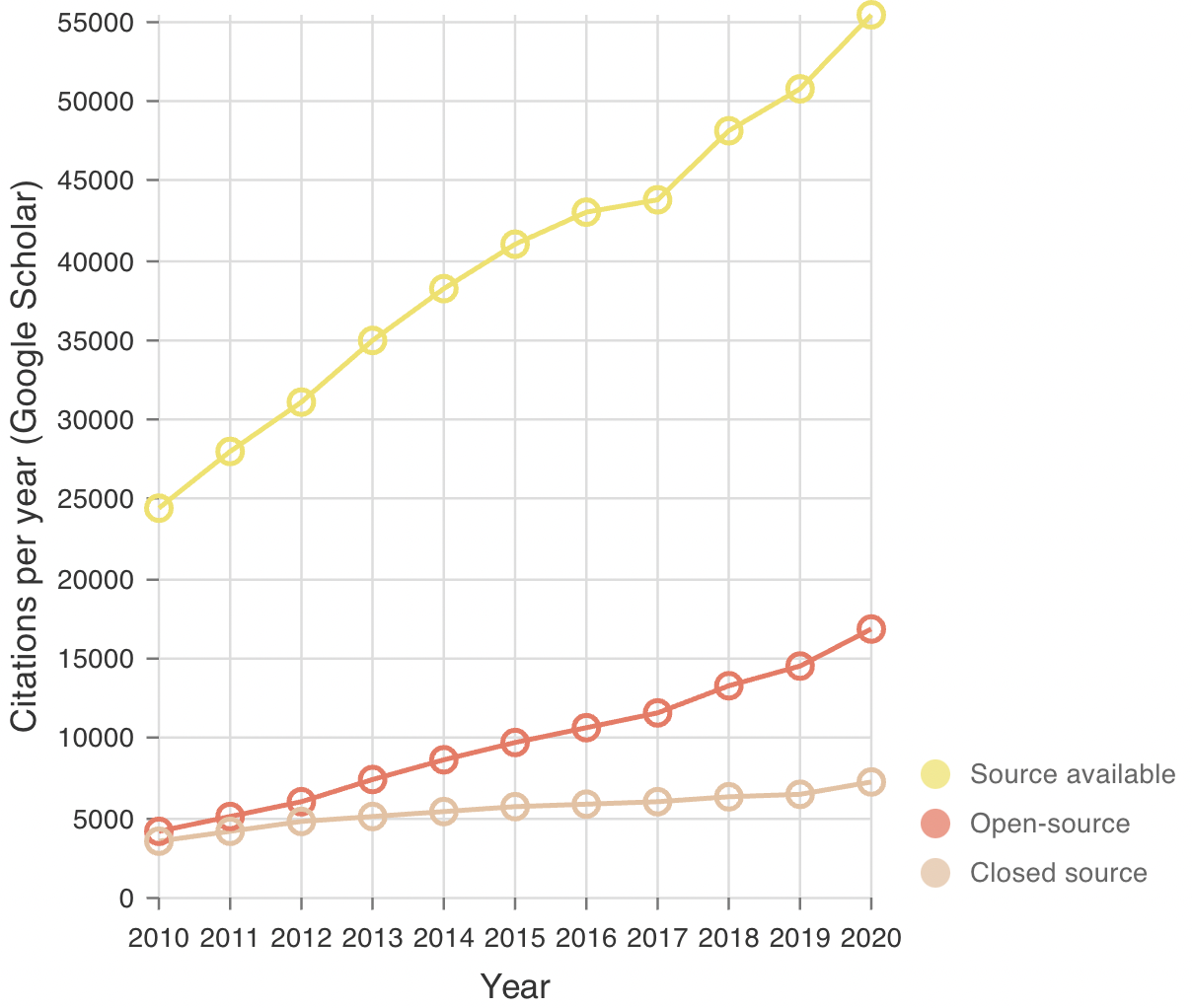}
    \caption{
        Citations by source code availability.
        "Source available" includes all engines whose source code can be obtained for free or for a fee.
        "Open-source" includes only OSI-approved licenses.
    }
    \label{fig:code-availability}
\end{figure}

Another important question concerns  source-code availability, which is relevant for the ability of researchers to independently verify published calculations and pin down bugs.
Figure~\ref{fig:code-availability} shows that consistently at least ${\sim}90\%$ of citations went to engines whose source code is available, strongly dominating over the 12 closed-source codes, whose citations have stagnated%
\footnote{One notable exception is the closed-source ORCA code \cite{Neese2020} that is free for academic use.} 
during the 2010s.
We recall here that counting citations in the scientific literature places a focus on the usage in academia, and that usage patterns in industry may differ from the trends identified here.

\begin{table}
    \centering
    \begin{tabular}{llll} \toprule
        License & Version & Copyleft & Published  \\ \midrule
        MIT &   &   & 1980s\\
        GPL & 2 & X & 1991\\
        LGPL & 2.1 & X & 1999\\
        BSD 3-clause & 2 &   & 1999\\
        Apache & 2 &   & 2004\\
        GPL & 3 & X & 2007\\
        LGPL & 3 & X & 2007\\
        ECL & 2 &   & 2006\\
\bottomrule
    \end{tabular}
    \caption{
        OSI-approved open-source licenses used in the collection.
        See the SPDX license list at \url{spdx.org/licenses} for the full license terms corresponding to the abbreviations.
    }
    \label{tab:open-source-licenses}
\end{table}

While citations to source-available engines have grown by ${\sim}130\%$ since 2010, 
citations to the 24 open-source engines within that group rose by >300\% within the same time frame, gaining market share.
In this context, it is useful to recall that the development of several engines on the lists predates the open-source movement and the creation of many of the open-source licenses that are in broad use today (see Table~\ref{tab:open-source-licenses}).
\atsoft distinguishes between 
\begin{itemize}
    \item 
        \emph{copyleft} open-source licenses, such as the GPL family, which require%
\footnote{Under specific circumstances, which differ significantly between the GPL and the LGPL.}
        derivative software to be distributed under the same open-source license (thus also called \emph{share-alike} or \emph{viral} licenses), and
\item
\emph{permissive} licenses, such as the BSD, Apache, and MIT licenses, which permit relicensing of derivative works.
\end{itemize}

The enforced sharing of improvements in derivative works can be a competitive advantage of adopting a viral license. 
It can also be one path towards financial revenue when companies seek a separate license agreement that allows them to keep derivative works proprietary.
Other developers may want to maximize impact of their software by lowering the barrier for adoption across the board, and thus prefer permissive licenses.
Overall, the choice of license is highly nuanced and an extensive discussion is beyond the scope of this article (interested readers are referred to \url{choosealicense.com}) but it may be instructive to observe the choices made by the codes in the collection.

Out of the 24 open-source codes in the \atsoft collection, the majority (20) adopt the GPL or LGPL license.
The four codes that are distributed under \emph{permissive} licenses (NWChem \cite{Apra2020}, OpenMM \cite{Eastman2017}, RASPA \cite{Dubbeldam2016a} and PySCF \cite{Sun2020a}) either switched to this licensing scheme in the late 2000s or 2010s or started being developed during that time.
This indicates that the use of permissive licenses is a recent phenomenon in the space of atomistic simulation engines, and may follow in the footsteps of the open-source community at large which is exhibiting a similar trend:
according to an analysis of over 4 million open-source packages by WhiteSource \cite{WhiteSource2021}, the use of permissive open-source licenses has nearly doubled from 41\% in 2012 to 76\% in 2020, with the Apache and MIT licenses alone accounting for more than half of all licenses that year.

So much about the differences between licensing models. 
Overall, citations of atomistic simulation engines in the collection have grown at an annual compound growth rate of ${\sim}8\%$, 
roughly twice the $4\%$ growth rate seen in the publication of peer-reviewed articles in science and engineering over the last decade \cite{citation-us}.
While part of this difference may reflect changing citation practises%
\footnote{According to Mammola et al., the length of reference lists in ecology journals has been increasing by ${\sim}2\%$ per year over the last two decades. \cite{Mammola2021}},
it likely indicates an increasing adoption of (atomistic) computational materials science throughout the scientific literature.

\section{Conclusions \& Outlook}

At the time of writing, the \atsoft collection contains over 60 simulation engines that each gather >100 citations per year, some several hundreds or thousands.
Overall, this review paints a bright future for the field of atomistic simulation: 
a growing variety of both commercial and free software to choose from, citation growth rates that substantially outpace the rest of the scientific literature, and a forecasted trillion dollar market potential for a digitally-driven materials revolution \cite{Satell2019}.
There are, however, some challenges ahead as well.

The continued slowdown in single-core performance scaling%
\footnote{While supercomputers have historically been able to escape this trend by driving up parallelization, Figure~\ref{fig:top500} suggests that the era of exponentially scaling supercomputer performance may also be coming to an end.}
creates a powerful driving force for the specialization of computer hardware.
Small- to medium-size development teams often lack the expertise or the resources to adapt their code base to an ever growing number of hardware accelerators and are at risk of falling behind.
One way of approaching this issue is to try and identify low-level, performance-critical primitives that are needed by multiple codes. 
These primitives can then be bundled into domain-specific libraries, such as libxc \cite{Lehtola2018}, libint \cite{Valeyev2021}, ELSI \cite{Yu2020}, SIRIUS \cite{Sirius2021}, M-A-D-N-E-S-S \cite{Harrison2016}, or TiledArray \cite{Calvin2021} that are ported to and optimized for the various accelerator architectures by HPC specialists.

Another issue that requires attention is the one of software citation.
With the increasing role that software plays in advancing science, it is crucial that credit for the creation of software is attributed adequately and accurately.
One reason why the citation counts in \atsoft are approximate is that software citation in the field of atomistic simulations comes in many different forms: 
\begin{itemize}
    \item references of papers that summarize recent developments of the software,
    \item references of papers that describe the implementation of specific methods,
    \item references to the home page of the code, or even just
    \item mentioning the code by name in the main text, possibly followed by a key author or company in parentheses,
\end{itemize}
similar to what has been found in the field of biology \cite{Howison2016}.
Furthermore, there is no standardized way of expressing whether a specific version of the software was used or whether the software was referenced as a general concept, e.g. as part of an enumeration of different codes like in this review.

In order to encourage adoption of a consistent policy for software citation across disciplines and venues, in 2016 the software citation working group of the FORCE11 coalition (\url{force11.org}) issued detailed software citation principles \cite{Smith2016}, which include the recommendation of citing a unique, persistent identifier that indicates which version of the software has been used.
Today, the technological infrastructure for \emph{creating} these identifiers is in place:
for example, for software hosted on \url{github.com}, the Zenodo-GitHub integration \cite{zenodo-github-integration} automatically stores the source code of each software release on the Zenodo repository operated by CERN, and mints a document object identifier (DOI) for it.

\begin{figure}
    \includegraphics[width=0.5\textwidth]{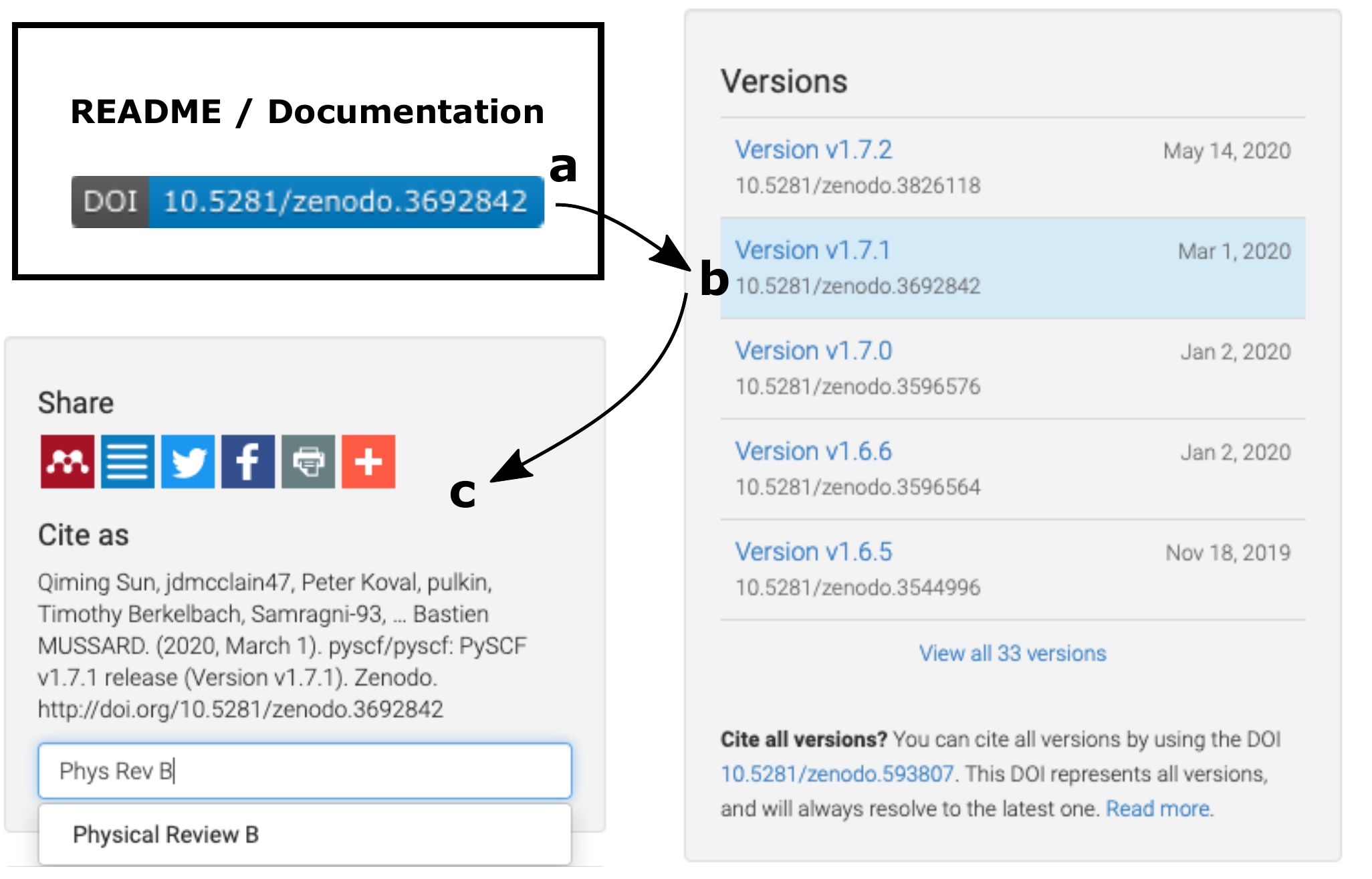}
    \caption{
        User flow for citation via Zenodo-GitHub integration.
        (a) User clicks on DOI badge in the README file of the source code repository or in the documentation of the software.
        (b) User is redirected to the landing page of the Zenodo software record, where they can pick the version they used.
        (c) User enters the desired citation style and copies the citation \cite{Sun2020c} into their manuscript or downloads the citation in a format supported by their reference manager.
    }
    \label{fig:zenodo-github}
\end{figure}

Figure~\ref{fig:zenodo-github} illustrates how citing such a DOI works from the user's perspective:
When code developers place the DOI badge offered by Zenodo in the "how to cite" section of their documentation, users can click on it and be redirected to the landing page of the accompanying Zenodo record.
There, they select the DOI corresponding to the version they used -- or, if they are referring to the software in general, they can cite the "concept DOI" of the software that represents all versions and always resolves to the latest one.
Finally, they can select the desired citation style and copy the citation into their manuscript or download the citation in a format supported by their reference manager.

At least three codes in the \atsoft collection (PySCF \cite{Sun2020a,Sun2020b}, OpenMM \cite{Eastman2017,peastman2020}, and xtb \cite{Bannwarth2021,Grimme2021}) have already enabled the Zenodo-GitHub integration but none of them mention this in their citation recommendations yet, effectively reducing the functionality of the integration to that of a future-proof backup of individual software versions.
This is a common theme found across software records on Zenodo today \cite{vandeSandt2019a}.
Part of the reason may be that Zenodo records are not (yet) indexed by the widely used scholarly search engines, such as the Web of Science, Scopus or Google Scholar.
But as researchers are getting increasingly accustomed to using platforms like Zenodo, the Open Science Framework (\href{https://osf.io}{osf.io}) or Figshare (\href{https://figshare.com}{figshare.com}) for depositing and citing \emph{data sets}, it seems to be just a matter of time until analogous practices in software citation will reach the mainstream.
Zenodo already \emph{does} track citations of its records through publicly available sources such as Crossref and Europe PubMedCentral (and displays them on the record). 
Trailblazing developers can therefore recommend their users to cite a version-specific Zenodo DOI in \emph{addition} to a review paper, and thereby get valuable statistics on which versions of their software are being used in return.
It can also be a convenient way of making a new code citable before a paper has been written on it.

As for the \atsoft collection, this review only marks the beginning.
Going forward, the collection will receive annual updates, including updates of this perpetual review when warranted.
Possible directions for further work include
\begin{itemize}
    \item 
adding any simulation engines that were missed, 
    \item 
recording the time evolution of licenses at the level of individual codes, and
    \item 
potentially evolving the scope of the collection, e.g. to include software for atomistic visualization or workflow management (although care would need to be taken in order not to lose focus).
\end{itemize}
Suggestions for future directions as well as updates and corrections of engine metadata (search keywords, tags, distribution channels, accelerator support, supported APIs, benchmarks, \ldots) are highly welcome,
be it through public discussions on the \atsoftgit issue tracker, via pull requests to the repository or via private communication to the authors.

\section{Author Contributions}
\textbf{Leopold Talirz:} Conceptualization, Methodology, Software and Data curation for atomistic.software. Writing- Original Draft, Reviewing and Editing.\\
\textbf{Luca M. Ghiringhelli:} Conceptualization, Methodology and Data curation for the original static version of the collection. Writing- Reviewing and Editing.\\
\textbf{Berend Smit:} Supervision, Writing- Reviewing and Editing, Funding acquisition

\section{Other Contributions}
The authors are grateful to Nicola Marzari and Matthias Scheffler for sharing their original idea of a list of atomistic simulation codes combined with citation data. 
Thanks also go to Nicola Marzari for providing the idea for table~\ref{tab:aps-most-cited} and to Kevin Jablonka for a careful reading of the manuscript.
Insightful discussions with Michele Ceriotti, Lars Holm Nielsen, Kevin Jablonka, Nicola Marzari, and Paul Saxe have contributed to the development of this work, as have numerous constructive suggestions by the anonymous reviewers of the manuscript.

For a more detailed description of contributions from the community and others, see the GitHub issue tracking and changelog at \githubrepository.

\section{Potentially Conflicting Interests}
The authors declare no competing interests.

\section{Funding Information}
This work was supported by the MARVEL National Centre for Competence in Research funded by the Swiss National Science Foundation (grant agreement ID~51NF40-182892), 
and the ``MaGic'' project of the European Research Council (grant agreement ID~666983).

\section*{Author Information}
\makeorcid

\printbibliography

\end{document}